\documentclass[journal]{IEEEtran}

\usepackage{cite}
\usepackage{units}
\usepackage[dvips]{graphicx}
\usepackage{amsthm}
\usepackage{enumerate}
\usepackage[cmex10]{amsmath}
\usepackage{amssymb}
\usepackage{fancyhdr}
\usepackage{framed}
\usepackage{algorithm,algorithmic}
\newtheorem{definition}{Definition}
\newtheorem{theorem}{Theorem}

\newtheorem{remark}{Remark}

\begin{document}

\title{A Ruin Theoretic Design Approach for Wireless Cellular Network Sharing with Facilities}
\author{Malcolm Egan$^{1}$, Gareth W.~Peters$^{2}$, Ido Nevat$^{3}$ and Iain B. Collings$^4$

\begin{center}
{\footnotesize {\
\textit{
$^{1}$ Faculty of Electrical Engineering, Czech Technical University in Prague, Czech Republic.\\
$^{2}$ Department of Statistical Sciences, University College London (UCL), London, England.\\
$^{3}$ Institute for Infocomm Research, Singapore.\\
$^{4}$ Department of Engineering, Macquarie University. \\[0pt]
} } }
\end{center}
}
\maketitle
\thispagestyle{empty}

\begin{abstract}

With the rise of cheap small-cells in wireless cellular networks, there are new opportunities for third party providers to service local regions via sharing arrangements with traditional operators. In fact, such arrangements are highly desirable for large facilities---such as stadiums, universities, and mines---as they already need to cover property costs, and often have fibre backhaul and efficient power infrastructure. In this paper, we propose a new network sharing arrangement between large facilities and traditional operators. Our facility network sharing arrangement consists of two aspects: leasing of core network access and spectrum from traditional operators; and service agreements with users. Importantly, our incorporation of a user service agreement into the arrangement means that resource allocation must account for financial as well as physical resource constraints. This introduces a new non-trivial dimension into wireless network resource allocation, which requires a new evaluation framework---the data rate is no longer the only main performance metric. Moreover, despite clear economic incentives to adopt network sharing for facilities, a business case is lacking. As such, we develop a general socio-technical evaluation framework based on ruin-theory, where the key metric for the sharing arrangement is the probability that the facility has less than zero revenue surplus. We then use our framework to evaluate our facility network sharing arrangement, which offers guidance for leasing and service agreement negotiations, as well as design of the wireless network architecture, taking into account network revenue streams.

\end{abstract}



\section{Introduction} \label{Introduction}

In traditional wireless networks, expensive infrastructure and the rapid adoption of new radio technologies resulted in small numbers of new market entrants. More recently, sharing of the radio access network (RAN) has been adopted to reduce capital expenditures (including infrastructure) of new operators, while still offering wide coverage and high quality of service (QoS). This trend has encouraged new operators to enter the market and sophisticated arrangements between infrastructure owners and operators are being considered.

There are many possible network sharing arrangements for wireless cellular network infrastructure, including mast sharing, full RAN sharing, roaming, or core-network sharing. At present, the most common sharing arrangements have involved only the RAN \cite{GSMA2012}, which allows operators to pool resources and can increase the capacity available for operators to service regions that already have high base station (BS) density or to improve coverage to under-served areas in developing regions. A common RAN sharing arrangement is based on companies that only offer base stations to operators (known as tower companies), which are now a large part of the wireless industry; particularly in India and the United States. In other regions, established operators have heavily invested in infrastructure and negotiated BS sharing arrangements with new entrants to the market, such as in Sweden. Ultimately, these approaches can be viewed as the first steps towards the dynamic market-based ``networks without borders'' vision where resources including spectrum, RANs and core networks are pooled, with contributors ranging from individuals to traditional operators \cite{Doyle2014}.

In parallel with the adoption of network sharing, small-cell technology (also known as femto, micro, pico, or metro-cells depending on the provider and transmitting capabilities) is revolutionizing the wireless industry with cheap, low-power base stations \cite{Ghosh2012}. By exploiting small-cells, wireless networks can offer high data rates with a small footprint via dense placement, which reduces the distance between the small-cell and the user---the most effective means of increasing network capacity \cite{Andrews2013}.

Despite the success of current network sharing arrangements and small-cells, there remain complex issues to be resolved. In particular, property must be leased to anchor small-cells, and backhaul must be installed in order to implement network MIMO (including CoMP) as well as inter-cell interference coordination (eICIC) \cite{Lopez-Perez2011}---key techniques for the effective operation of small-cell networks. Moreover, operational expenditures (OpEx)---largely due to small-cell maintenance---are growing with the increasing number of small-cells in the network. This increase in OpEx will continue until effective self-organizing network (SON) technologies \cite{ElSawy2013a} are successfully implemented.

A particularly challenging scenario for operators is large facilities with high data rate demands, even when standard network sharing arrangements are employed. Common examples of these facilities are universities, stadiums, convention centers, utilities, mines, and high density urban residences. The challenge arises due to the high cost in leasing small-cell locations, cost of leasing high bandwidth backhaul (despite the fact that the backhaul is often available within these facilities via fibre links), the often unusual characteristics of user demands---such as high upload rates in stadiums \cite{Calin2013}---and large variations in data rate demand over time. Moreover, large facilities may desire to only charge low rates to users in order to improve the operation of the facility; in stark contrast with standard wireless service arrangements. This can occur either because the users are employees and the mobile service is paid for by the facility, or a cheap mobile service is used as an attraction to the facility; similar to how WiFi is currently offered free of charge to users in many convention centers and hotels.

\subsection{A New Socio-Technical Network Sharing Approach}

There is strong motivation for large facilities to offer cheap mobile services and to develop their own capability to do so; namely, the low cost of small-cells, and the availability of high bandwidth backhaul. As such, there is a genuine need for an alternative network sharing arrangement that can exploit the unique characteristics of large facilities.

In this paper, we propose a new network sharing arrangement for facilities. Our facility network sharing arrangement is based on a network consisting of small-cells operated by and within the facility, which contrasts with tower companies that own multiple small-cells and BSs, which are leased to operators.

There are two key aspects to our facility network sharing arrangement: the core network leasing agreement between traditional operators and the facility; and the service agreement between the facility and users. The purpose of the core network leasing agreement is to provide the facility with data to serve users that are subscribed to traditional operators as well as spectrum localized to the facility. This agreement involves a fee that the facility pays to the traditional operator, in exchange for core network access and spectrum. On the other hand, the service agreement is the mechanism by which the facility obtains revenue. In particular, the facility charges users for its services; depending on the resources required to ensure reliable transmission, and also pricing parameters designed to compensate for the leasing fee for core network access.

In concept, our facility network sharing arrangement bears similarities to the ``local network operated by an independent actor'' classification for indoor cellular networks proposed in \cite{Markendahl2013}. More specifically, the approach in \cite{Markendahl2013} introduced the notion of third party operators to service users inside large buildings, which negotiate with traditional operators for data access. Our facility network sharing arrangement differs in two key aspects: (i) we are not limited to indoor operation, which is achieved using a general stochastic geometry model that can in principle be extended to include small-cell cooperation via eICIC techniques; and (ii) we propose a specific agreement structure between operators, users, and facilities. In particular, core network access is provided via a new leasing arrangement that can in principle be formulated as a contractual obligation, which we detail in Section~\ref{sec:leasing}.

While facilities have currently not been considered within a network sharing arrangement, there are six key economic incentives, subject to the presence of appropriate antitrust regulation: (1) the facility already owns or leases the property, which means that there is zero sunk cost for positioning small-cells; (2) the facility's high bandwidth fibre network can be exploited at no additional cost to provide high rates through advanced eICIC; (3) competition will be increased between traditional operators in the region through core network access costs (in contrast with RAN discrimination); (4) users will be charged only at the facility's incremental costs as the facility is either covering the costs itself (e.g., in mines or utilities) or the service is offered as an attraction (e.g. in hotels or convention centers); (5) facilities can offer more efficient power sources as they also must power the facility (e.g., through large-scale solar cell sources); and (6) SON innovation is promoted as the facilities chase a reduced OpEx. An incremental approach to SON is also possible since the facilities will have a relatively small number of small-cells, which means that there are fewer complexities compared with the large-scale RANs of traditional operators.

These benefits clearly suggest that there is a significant potential for network sharing arrangements involving facilities.

\subsection{Evaluation of Facility Network Sharing Arrangements}

Although there are clear economic incentives for the adoption of facility network sharing arrangements, a business case based on a quantitative framework is lacking. We address this issue by evaluating facility network sharing using a new framework, which incorporates both the wireless communications network, as well as the service and leasing agreements between users, the facility, and traditional wireless operators that own the core networks. In contrast with standard analytical frameworks in the wireless communications literature, our framework models the facility as a \textit{socio-technical system}. This means that the conclusions arising from our framework lie close to those used directly in real-world practice. Moreover, our framework is general, which means that it can be adapted to a range of wireless settings, where resource sharing is used and financial sustainability is a key design criterion.

Profit-based approaches for resource allocation in cellular networks have been considered in \cite{Duan2011,Park2014,Berry2010,Berry2013,Yang2014,Halluin2002}. Early work on capacity pricing adapted a real options framework to cope with uncertainties \cite{Halluin2002}. In \cite{Duan2011,Yang2014}, cognitive cellular networks and small-cell networks, respectively, were modeled via Stackelberg games. The approach in \cite{Yang2014} employed subsidies to incentivize closed access small-cell access points to share allocated spectrum. The approaches in \cite{Berry2010,Berry2013} also exclusively focused on spectrum allocation. In \cite{Park2014} the BS density and spectrum usage were optimized to maximize the net profit; however, this analysis was based solely on long-term average rates. The drawback of such an approach is that it does not consider any assessment of the feasibility of the business model nor optimal solutions obtained from the perspective of a financial risk management analysis. As such, it is not easy to assess the long-term profitability of the operator. This is important as network design that takes profitability into account will differ from standard designs under purely technical constraints and generally lead to improved operator longevity. Key parameters such as compounded interest rates, initial capital, expenditures, and future investments also cannot be incorporated into the standard approaches. Moreover, previous work does not account for the effect of fluctuations in the number of users and their cellular data demands.

Jointly addressing a number of the key financial and technical factors that arise in real-world cellular networks (including features such as channel gains, power control, path loss and the duration of user connections), is crucial to reliably predicting and optimizing the financial sustainability of network sharing arrangements. However, this requires a more sophisticated socio-technical framework. In this paper, we propose a new quantitative framework to evaluate facility network sharing arrangements based on a reformulation of ruin theory$^1$ \cite{Asmussen2010} to facilitate a study of business models, where the key performance metric is the probability that the owner of the facility has a negative revenue surplus within a period of $n$ months---leading to financial insolvency. We emphasize that the probability of these events, termed ruin, is dependent on both economic and technical factors, which include: the initial capital of the facility; the compound interest rate (compounded monthly); link parameters, such as channel gains, power control and path loss; duration of user connections; and the number of users in the network.

\footnotetext[1]{
Historically, ruin theory originally arose in the study of solvency requirements for insurance \cite{Asmussen2010}.
}

\subsection{Key Contributions}

We summarize our three key contributions as follows.
\begin{enumerate}
\item We propose a new network sharing arrangement between traditional operators, facilities, and users. The arrangements accounts for both the expenditures and revenue for the facility, which arise due to the cost of leasing access to traditional operators' core networks (expenditures), and the income from servicing users normally subscribed to traditional operators (revenue). Importantly, the service charges incurred by each user depend on the physical resources required to ensure reliable transmission. As such, the service charges are strongly coupled to the capabilities of the RAN.
\item We develop a quantitative ruin theory-based framework to evaluate facility network sharing. The evaluation is based on the probability of ruin; that is, the probability that there is a negative revenue surplus within a period of $n$ months (a standard metric for financial risk analysis \cite{Asmussen2010}). In order to obtain the probability of ruin, we:
    \begin{enumerate}
    \item Derive the moments of the revenue from user service charges, based on a practical wireless heterogeneous network model. In particular, we exploit stochastic geometry techniques \cite{ElSawy2013b} to model the placement of base stations and users.
    \item Use the moments of the revenue time-varying stochastic process to compute the probability distribution of the net profit derived by the facility operator by incorporating revenue from the service of each user.
    \item Compute the probability of ruin via the probability distribution of the net profit and efficient recursions (based on linear difference equations), which are motivated by results in insurance theory \cite{Hurlimann1991,Sun2003}. Importantly, we extend the applicability of the standard recursions account for idiosyncratic (from the perspective of insurance) aspects of wireless communication networks.
    \end{enumerate}
\item We evaluate facility network sharing based on numerical evaluation of the probability of ruin via our framework. First, we demonstrate that facility network sharing can be profitable. Second, we provide insights to guide the design of wireless networks to account for revenue and expenditures, in addition to ensuring reliable transmission.
\end{enumerate}

\section{Wireless Network Model}\label{sec:wireless_mod}

In this section, we detail the wireless networking aspects of our model for the facility (financial aspects will be discussed in Section~\ref{sec:leasing}). The key aspects that we consider are the placement of the small-cells, which small-cell a given user will connect to, and the physical layer transmission model. We note that our modeling assumptions are based on previous approaches to wireless cellular networks (see e.g., \cite{Andrews2011}) and the only significant difference is that the scale of facilities are typically smaller than those considered for standard networks.

Our assumptions for small-cell placement and the user selection protocol are as follows:
\begin{enumerate}
\item[\textbf{A1:}] The facility's small-cells are arranged according to a homogeneous spatial Poisson point process (PPP) with intensity $\beta$. This means that in each realization there is a different number of small-cells, which provides insight into the behavior of the network irrespective of how the small-cells are positioned.
\item[\textbf{A2:}] All users serviced by the facility:
\begin{enumerate}
\item[(i)] connect to the nearest facility small-cell;
\item[(ii)] are arranged according to an independent stationary point process, not necessarily Poisson.
\end{enumerate}
\end{enumerate}
The key consequence of assumptions (\textbf{A1}) and (\textbf{A2}) is that the distribution of the distance $R$ of a small-cell and any user it services is given by \cite[Section III-A]{Andrews2011}
\begin{align}\label{eq:U_BS_dist}
\mathrm{Pr}(Z \leq z) = e^{-\beta \pi z^2}2\pi z \beta.
\end{align}

We now detail our assumptions for the physical layer transmission model.
\begin{enumerate}
\item[\textbf{A3:}] The facility's small-cell network operates in discrete time with block fading. Each time slot (also corresponding to a fading block), has a duration (coherence time) of $T$ seconds.
\item[\textbf{A4:}] Each small-cell interferes with the others. Due to the Poisson interference assumption (\textbf{A1}), the interference is $M/M$ shot noise \cite{Andrews2011}. As such, the received signal in the $l$-th time slot ($l = 1,2,\ldots$) of a user's connection is given by
    \begin{align}
    y_l = h_l\sqrt{r_U^{-\alpha}P_0}x_l + n_l + z_l,
    \end{align}
    where $h_l \sim \mathcal{CN}(0,1)$ are the independent fading coefficients for time slots $l$, and $r_U$ is the distance between the user and its small-cell (the closest one, by (\textbf{A2})), distributed according to (\ref{eq:U_BS_dist}). $P_0$ is the power level the small-cell transmits at; not necessarily constant. $\alpha$ is the path loss exponent, $x_l \sim \mathcal{CN}(0,1)$ is the (Gaussian) data symbol, and $n_l \sim \mathcal{CN}(0,\sigma^2)$ is additive white Gaussian noise (AWGN), with noise variance $\sigma^2$. $z_l$ is the $M/M$ shot noise.
\end{enumerate}
Based on assumption (\textbf{A4}), the instantaneous signal-to-interference and noise ratio (SINR) in the $l$-th time slot of a given user's connection is given by
\begin{align}\label{eq:SINR}
\gamma_l = \frac{|h_l|^2r_U^{-\alpha}P_0}{\sigma^2 + I_{l}},
\end{align}
where $I_l$ is the interference power in the $l$-th time slot of the connection. In our analysis, we focus on the interference limited scenario where $\sigma^2 = 0$.

The achievable data rate in each time slot under Gaussian signaling for a user in the $l$-th slot of her connection is given by
\begin{align}\label{eq:rmax_l}
R^{\max}_l = B\log_2(1 + \gamma_{l}),
\end{align}
where $B$ is the bandwidth (a constant) and $\gamma_l$ is given by (\ref{eq:SINR}).

\section{Proposed Network Sharing Arrangement}\label{sec:leasing}

In this section, we introduce our network sharing arrangement between the facility, users, and traditional operators (which have their own core networks). It is important to consider traditional operators as their core networks are a key source of data for the facility's small-cell network as well as spectrum. Although it is also possible to obtain data from the internet (although not necessarily calls), when there are a significant number of users this can consume a large amount of bandwidth and may also require special agreements with ISPs. As such, we focus on the scenario where traditional operators provide data via their core networks.

\begin{figure}[!h]
\centerline{\includegraphics[height=2.3in]{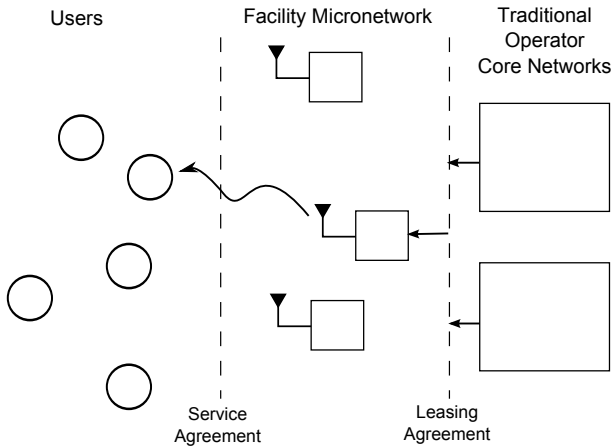}}
\caption{Network sharing arrangement diagram.} \label{fig:network_sharing}
\end{figure}

Our network sharing arrangement consists of two components: a leasing agreement with traditional operators; and a service agreement with users, which is illustrated in Fig.~\ref{fig:network_sharing}. The agreements determine financial exchanges (contractual obligations) between the facility, users, and the traditional operators. Moreover, these exchanges induce revenue and expenditure processes that are stochastic due to uncertainties in the resources required to provide the service (due to the stochastic nature of the wireless channel) and the random duration of the users' connections. To this end, we formalize these processes which forms the basis of our evaluation framework in Section~\ref{sec:eval}.

\subsection{Leasing and Service Agreements}

A leasing agreement between the facility and a traditional operator allows the facility to access the data required by users it seeks to service. In the leasing agreement, the facility pays a fee to the traditional operator in return for access to their core network and spectrum localized to the facility. A key feature of our leasing agreement is that the facility pays a fee \textit{per connection}. This means that the facility only pays for access it actually requires, which is consistent with the broader vision of dynamic wireless network sharing proposed in \cite{Doyle2014}.

We assume that the fee is only determined by the traditional operator that is providing the service; that is, it is independent of other factors such as the duration of the connection. As users are typically subscribed to a single operator, this means that the fee for each user may be different. Our leasing agreement is summarized as follows.
\begin{framed}
Our proposed \textit{core network leasing agreement} between traditional operators and the facility consists of the connection fee for the $j$-th user. The connection fee is given by
\begin{align}\label{eq:core_net_fee}
C(\tau_{i,j},k_{i,j}) = C^*(k_{i,j}),
\end{align}
where $C^*(k_{i,j})$ is the connection fee for users subscribed to operator $k_{i,j}$.
\end{framed}

Observe that the leasing agreement does not depend on the duration of each connection. As each core network typically has a large number of connections at any given time, it is likely that the distribution of the number and duration of each connection is known to the corresponding traditional operator. Moreover, the infrastructure required to support a given connection is also known due to the stability of the wired links (as opposed to the wireless scenario). In this case, the operator can obtain good estimates of equilibrium prices required to ensure that profit targets are met, which means that the duration of each connection does not play a key role. However, when there is a scarcity of access to the core network then it is necessary to consider connection durations in the fee, which may be achieved via a market mechanism such as an auction.

We now consider the service agreement between the facility and each user. The basis of the service agreement is that there is a cost to the facility to provide users with the wireless service. The cost depends on four key factors: (i) the traditional wireless operator that the user has a subscription; (ii) the duration of the users connection to the facility; (iii) the quality of the wireless service requested (i.e., the rate at which the user requests to be serviced); and (iv) the SINR of the link between the user and the facility small-cell that it is serviced by. \textit{Importantly, all of these factors are random quantities.}

The service agreement corresponds to the revenue the facility receives in exchange for servicing each user, which can be obtained from two sources. The first source is directly from the users; that is, the users pay the facility for the use of the service. In this case, the users pay the facility for every connection they require. The second source is the facility itself. In this case, the facility subsidizes the users and the agreement corresponds to the funds the facility allocates to cover the cost of servicing each user, which is applicable when the facility is providing the service for its employees or when the service is an attraction (e.g., for hotels or stadiums). Again, this is on a per connection basis, which means that users (or the facility) only pay for the connections that they actually use.

We assume that the service provider offers $Q$ products of varying QoS. In practice, the user selects an application that it seeks to use---such as voice, video, or data---and the service provider offers a QoS product that can support the application.

\begin{remark}[User indexing notation]
In a given time slot, multiple users are likely to conclude their connection. To refer to a given user, we index it by the pair $(i,j)$, where $i$ is the time slot that the user \textit{concludes} its connection and $j$ is the index of the user in the set of users that conclude their connection in time slot $i$. For example, when a user is the $4$-th user that terminates its connection in time slot $5$, then the user is indexed as $(5,4)$.
\end{remark}

In order to provide increasing levels of QoS over the wireless link, the facility is required to employ an increasing number of physical resources; for instance, additional bandwidth, power, or infrastructure (e.g. relays or distributed antennas). It is important to note that a different number of physical resources are usually required in each time slot to account for channel fading. As such, the facility varies the SINR in each time slot by a scaling factor $c_{i,j,l}$ (corresponding to the $l$-th time slot of the connection for the $(i,j)$-th user) to achieve
\begin{align}\label{eq:rate_prod}
R^{(q)}_{i,j} = B\log\left(1 + \gamma_{i,j,l}c_{i,j,l}\right),
\end{align}
where $R_{i,j}^{(q)}$ corresponds to the rate required to support QoS product $q \in \{1,2,\ldots,Q\}$ for user $(i,j)$ and $c_{i,j,l}$ encapsulates the additional physical resources required to support QoS product $q$. Observe that (\ref{eq:rate_prod}) follows directly from (\ref{eq:rmax_l}) with $\gamma_l$ replaced with $\gamma_{i,j,l}$, which clarifies the particular user that is serviced. While it might seem more natural to directly scale the rate (i.e. define $R^{(q)} = Bc'_{i,j,l}\log(1 + \gamma_{i,j,l})$), it is in fact simpler to use our formulation. We will show this in Section~\ref{sec:eval}, where we develop our evaluation framework based on ruin theory. It is also worthwhile to note that the two formulations are identical with the appropriate definition of $c'_{i,j,l}$.

In practice, the facility's physical resources are limited. As such, the scaling factor $c_{i,j,l}$ is upper bounded by a constant. We also introduce a lower bound on $c_{i,j,l}$ so that the facility is guaranteed a minimum income, even when the channel is good. This means that facility can ensure that it can pay the core network leasing fee, detailed in (\ref{eq:core_net_fee}). Taking these considerations into account, we define $c_{i,j,l}$ as
\begin{align}\label{eq:scale_fact}
c_{i,j,l} = \max\left\{\min\left\{\frac{2^{R^{(q)}_{i,j}/B} - 1}{\gamma_{i,j,l}},c_{\max}\right\},c_{\min}\right\}.
\end{align}

It is also necessary to account for the duration of each user's connection. In particular, we make the following assumptions:
\begin{enumerate}
\item[\textbf{A5:}] Let $\tau_{i,j}$ be the duration (in integer time slots) of the connection of the $j$-th user to end its connection in the $i$-th time slot; i.e user $(i,j)$. Each duration $\tau_{i,j}$ is independent and identically distributed. The durations $\tau_{i,j}$ has CDF $F_{\tau_{i,j}}$ with support $\{1,2,\ldots,i\}$.
\end{enumerate}

We now detail our proposed service agreement.

\begin{framed}
Our proposed \textit{service agreement} consists of the charge to the $j$-th user to end its connection in the $i$-th time slot (user $(i,j$)), which is given by
\begin{align}\label{eq:core_net_fee}
v(\tau_{i,j},\mathbf{c}_{i,j}) = \sum_{l=1}^{\tau_{i,j}} c_{i,j,l} T \rho,
\end{align}
where $\mathbf{c}_{i,j} = [c_{i,j,1},\ldots,c_{i,j,\tau_{i,j}}]^T$ is the vector of scaling factors (each element corresponding to a different time slot) to satisfy (\ref{eq:rate_prod}) and $\rho$ is the premium rate; that is, the income received from user $(i,j)$ by the facility per time slot with a unit scaling factor. We assume that the premium rate $\rho$ is constant, irrespective of the operator that user $(i,j)$ is subscribed to.
\end{framed}

\subsection{Summary of the Proposed Network Sharing Arrangement}\label{sec:proposal_summary}

From the perspective of the facility, the leasing agreement corresponds to an expenditure process and the service agreement corresponds to an income process. As such, our network sharing arrangement can be formalized as a revenue surplus process, which is defined in terms of the expenditure and income processes.

An important feature of the revenue surplus process is that interest is compounded. This means that the total revenue surplus is dependent on not only the current surplus, revenue and costs, but also the interest rate. Moreover, the time interval between interest compounding is typically different to the duration of a time slot. The number of users with connections ending in compound interest interval $m$ is given in (\textbf{A6}).
\begin{enumerate}
\item[\textbf{A6:}] The number of users $N_n$ that have their connection end in compound interest interval $[n-1,n)$ of duration $\kappa T$ ($\kappa \in \mathbb{N}$) is geometrically distributed as
\begin{align}
\mathrm{Pr}(N_i = u) = (1 - w_N)^uw_N.
\end{align}
\end{enumerate}

We first formalize the facility expenditure process, which is based on the core network leasing agreement with the traditional operators.
\begin{definition}[Facility Expenditure Process]\label{def:expend_proc}
Consider the $(i,j)$-th user associated with the $k$-th wireless operator, which has:
\begin{enumerate}
\item[(i)] a random connection duration $\tau_{i,j}$ time intervals, distributed according to $\tau_{i,j} \sim F_{\tau_{i,j}}$ (detailed in (\textbf{A6}));
\item[(ii)] a requested rate product $R_{i,j}^{(q)}$ with $q \in \{1,2,\ldots,Q\}$, distributed according to $R_{i,j}^{(q)} \sim F_{R_{i,j}}$ for any discrete distribution on $\{1,2,\ldots,Q\}$;
\item[(iii)] and required resources $\mathbf{c}_{i,j} = [c_{i,j,1},\ldots,c_{i,j,\tau_{i,j}}]^T$, which are i.i.d random variables defined in (\ref{eq:scale_fact}).
\end{enumerate}
Then the total expenditure of the facility in the compound interest interval $[n-1,n)$ is given by
\begin{align}
E_n = \sum_{j=1}^{N_n} C(\tau_{m,j},k_{m,j}),
\end{align}
where $(n-1)\kappa T \leq m < n \kappa T$ are the time slots that end in interval $[n-1,n)$, and $N_n$ is geometrically distributed (as detailed in (\textbf{A6})) and $C(\tau_{n,j},k_{n,j})$ is defined in (\ref{eq:core_net_fee}).
\end{definition}

Next, we formalize the facility income process, which is based on the service agreement with the users.
\begin{definition}[Facility Income Process]\label{def:income_proc}
Consider the $(i,j)$-th user associated with the $k$-th wireless operator, which satisfies the hypotheses in Definition~\ref{def:expend_proc}. Then, the proposed charge to the $(i,j)$-th user (corresponding to the facility income) for connecting to the facility for a duration $\tau_{i,j}$ is given by the compound random sum
\begin{align}
v(\tau_{i,j},\mathbf{c}_{i,j}) = \sum_{l=1}^{\tau_{i,j}} c_{i,j,l} T\rho.
\end{align}
The total income generated in the period $[n-1,n)$ is given by
\begin{align}
I_n = \sum_{j=1}^{N_n} v(\tau_{m,j},\mathbf{c}_{m,j}),
\end{align}
where $(n-1)\kappa T \leq m < n \kappa T$ are the time slots that end in interval $[n-1,n)$, and $N_n$ is geometrically distributed, as detailed in (\textbf{A7}).
\end{definition}

Finally, the revenue surplus process for the facility is defined as follows. This can be viewed as the accumulation of the initial capital and the difference between the income and expenditure processes, taking into account compound interest.
\begin{definition}\label{def:rev_surp}
The revenue surplus is the current micro-network profit in the $l$-th time slot generated by serving users, after accounting for interest and the cost of accessing the operators' core networks. This is given by
\begin{align}
S_l(u) &= u(1 + r)^l + \sum_{i=1}^l (1 + r)^{l-i} \sum_{j=1}^{N_i}\bigl[v(\tau_{m_i,j},\mathbf{c}_{m_i,j})\bigr.\notag\\
&~~~\bigl.- C(\tau_{m_i,j},k_{m_i,j})\bigr],
\end{align}
where $i \kappa T \leq m_i < (i+1) \kappa T$, $u$ is the facility's initial capital and $r$ is the compound interest rate (compounded at intervals of $\kappa T$). We also define
\begin{align}
S_{net}(i)= \sum_{j=1}^{N_i}\left[v(\tau_{m,j},\mathbf{c}_{m,j}) - C(\tau_{m,j},k_{m,j})\right],
\end{align}
as the net profit in compound interval $i$, where $i \kappa T \leq m < (i+1) \kappa T$.
\end{definition}

It is important to note that there is potentially a significant difference in the time scales at which users are priced (order of minutes) and that the revenue surplus is calculated (order of months). 

\section{Evaluation Framework: A Ruin Theory Approach}\label{sec:eval}

In this section, we detail our framework to evaluate our facility network sharing arrangement, which will be evaluated in Section~\ref{sec:sim_results}. Our framework is based on ruin theory \cite{Asmussen2010}, where the key performance metric is the probability that the facility has a negative revenue surplus within $n$ months, known as the probability of ruin. This is an important metric as the facility can only operate while it has the resources to do so. In fact, knowing whether these financial resources are likely to be available is important in the decision of whether or not to invest in the facility, how to structure products, and how much to charge for services.

Unfortunately, it is not straightforward to directly compute the probability of ruin. The main reason for this is that the income and expenditure processes (defined in Section~\ref{sec:proposal_summary}) involve a number of random variables, which result in random sums without closed-form distributions. Due to these difficulties, we instead focus on obtaining an accurate approximation of the probability of ruin.

Fortunately, we are able to leverage techniques from ruin theory to compute an accurate approximation for the probability of facility ruin. However, it is important to note that modifications to the standard theory are required, due to the fact that the parameters of facilities yield non-standard distributions.

In this section, there are four subsections: (A) the definition and overview of the key steps to compute the facility ruin probability; (B) the first step: the approximation of the income distribution, which is based on an orthogonal polynomial representation; (C) the second step: the recursion for the net profit distribution; and (D) The third step: the recursion for the probability of ruin.

\subsection{Ruin Probability Definition and Overview}

Intuitively, the probability of ruin is the probability that the revenue surplus is negative before a period of $l$ time periods (e.g., months). First, we define the stopping time known as the time of ruin, followed by the ruin probability.
\begin{definition}[Ruin Time]
The time of ruin is the first time that the revenue surplus is negative, i.e.
\begin{align}
L_R = \inf\{l: S_l(u) < 0\},
\end{align}
where $u$ is the initial capital. Note that we allow for the possibility that $u < 0$.
\end{definition}

\begin{definition}[Probability of Ruin]\label{def:ruin_prob}
The probability of ruin before a period of $l$ time periods is then
\begin{align}
\psi_l(u) = \mathrm{Pr}(L_R \leq l),
\end{align}
and the probability of survival is
\begin{align}
\phi_l(u) = 1 - \psi_l(u).
\end{align}
\end{definition}

In order to compute the ruin probability; the steps are detailed in Algorithm~\ref{alg:Ruin_Prob_Calc}.
\begin{algorithm}
\caption{Ruin Probability Computation}
\label{alg:Ruin_Prob_Calc}
\begin{algorithmic}
\item[1.] Derive the orthogonal polynomial basis expansion for the income PDF in Section~\ref{sec:orthog_rep}.
\item[2.] Evaluate the distribution for the net profit in time slot $l$ in Section~\ref{sec:panjer_rec}. This is achieved by first discretizing the income PDF from Section~\ref{sec:orthog_rep}, then using a recursion from actuarial science to compute the compound distribution.
\item[3.] Evaluate the probability of ruin via another recursion in Section~\ref{sec:ruin_rec} (different to the recursion in Stage 2), which circumvents the difficulty of directly computing the probability of ruin.
\end{algorithmic}
\end{algorithm}

\subsection{Orthogonal Polynomial Representation of the Income PDF}\label{sec:orthog_rep}

The first stage of computing the probability of ruin is to compute the PDF of the income from a user with connection ending in time slot $i$, which from Definition~\ref{def:income_proc} is given by
\begin{align}\label{eq:V_i}
V_i = v(\tau_{i,j},\mathbf{c}_{i,j}) = \sum_{l=1}^{\tau_{i,j}} c_{i,j,l} T\rho,
\end{align}
which is a random sum due to $\tau_{i,j}$ and $c_{i,j,l}$ (given in (\ref{eq:scale_fact})).
\begin{remark}\label{rem:dist_vary}
In general, the distribution of $\tau_{i,j}$ depends on $i$. For short compounding intervals this is to ensure that the connection duration is not longer than the time the system has been running. In other cases, the distribution may vary due to seasonal usage trends and events, such as holidays. This has the important consequence that the moments and hence distribution of $V_i$ depend on the time slot $i$ that user $(i,j)$'s connection ends.
\end{remark}
It is clear from the fact that $V_i$ is a random sum and the form of $c_{i,j,l}$ (see (\ref{eq:scale_fact})) that the distribution of $V_i$ is not readily obtained in closed-form. As such, we instead adopt a principled approach to approximating $V_i$, which is based on an Askey-orthogonal polynomial expansion. In particular, we use the Jacobi polynomials since the support of the income $V_i$ is bounded on $[-v_{i,\min},v_{i,\max}]$. This follows from the fact that both $\tau_{i,j}$ and $c_{i,j,l}$ are bounded.

It is important to note that the Jacobi polynomials are only orthogonal on $[-1,1]$. Hence, we need to transform $V_i$ so that it has also has support $[-1,1]$. This is achieved via the transformation
\begin{align}\label{eq:W_transform}
W_i = \frac{2(V_i + v_{i,\min})}{v_{i,\max} + v_{i,\min}} - 1,
\end{align}
where $v_{i,\min} = -c_{\min}T\rho$ and $v_{i,\max} = ic_{\max}T\rho$, from (\ref{eq:V_i}).

The distribution of $W_i$ via the Jacobi polynomial representation is then given by \cite{Cruz2014}
\begin{align}\label{eq:W_exp}
f_{W_i}(x) \approx K(x)\sum_{m=0}^d a_mP^{(a,b)}_m(x),
\end{align}
where $d$ is the order of the approximation, $P^{(a,b)}_m(x)$ is the $m$-th Jacobi poynomial with parameters $a,b$, $a_m$ is given by
\begin{align}
a_n &= \frac{B(a,b)(2n + a + b - 1)\Gamma(n + a + b - 1)n!}{\Gamma(n + a)\Gamma(n + b)}\notag\\
&~~~\times \int_{-1}^1 f_{W_i}(x)P_n^{(a,b)}(x)dx\notag\\
&= b_n \int_{-1}^1 f_{W_i}(x)P_n^{(a,b)}(x)dx,
\end{align}
and $K(x)$ is given by
\begin{align}
K(x) = \frac{(1 + x)^{a-1}(1 - x)^{b - 1}}{B(a,b)2^{a + b - 1}},
\end{align}
where
\begin{align}
B(x,y) = \Gamma(x)\Gamma(y)/\Gamma(x + y).
\end{align}
We note that the coefficients $a_n$ minimize the mean square error between the approximation and $f_{W_i}(x)$ as $d \rightarrow \infty$.

\begin{remark}
From extensive numerical experiments, we have found that the choice $a = b = 1$ (corresponding to Legendre polynomials) yields the most accurate approximations for a range of moments (corresponding to different network setups).
\end{remark}

As $P_n^{(a,b)}$ is a polynomial, we can write $a_n$ as
\begin{align}\label{eq:an_ev}
a_n &= b_n\int_{-1}^1 f_{W_i}(x) \sum_{s=0}^n \zeta_{n,s} x^{s} = b_n\sum_{s=0}^n \zeta_{n,s} \mathbb{E}[W_i^s],
\end{align}
where $\zeta_{n,s}$ corresponds to the coefficient of the $s$-th order term of $P_n^{(a,b)}$. This means that the approximation is completely characterized by the raw moments of $W_i$. Observe that the moments of $W_i$ are related to the moments of $V_i$ via
\begin{align}
\mathbb{E}[W_i^s] = \mathbb{E}\left[\left(\frac{2(V_i + v_{i,\min})}{v_{i,\max} + v_{i,\min}} - 1\right)^s\right],
\end{align}
which can be readily evaluated (given the moments of $V_i$) via the binomial expansion.

\begin{remark}
To obtain the moments of $V_i$ (and hence the moments of $W_i$), we compute and then differentiate the moment generating function, which allows us to use the probability generating functional of the PPP; surpassing the need to explicitly derive the distribution of the interference. The moments and details of the derivations are given in Appendix~\ref{app:moments}.
\end{remark}

The distribution of $V_i$ is then obtained from the distribution of $W_i$ via the transformation
\begin{align}\label{eq:VW_pdf_trans}
f_{V_i}(x) = \frac{2}{v_{i,\max} + v_{i,\min}}f_{W_i}\left(\frac{2(x + v_{i,\min})}{v_{i,\max} + v_{i,\min}} - 1\right),
\end{align}
where $f_{W_i}(x)$ is given by (\ref{eq:W_exp}). We summarize the procedure in Algorithm~\ref{alg:basis_exp}.

\begin{algorithm}
\caption{Orthogonal Polynomial Basis Expansion of the Net Profit PDF}
\label{alg:basis_exp}
\begin{algorithmic}
\item[1.] Compute the moments of $V_i$ using (\ref{eq:cond_moments}) and (\ref{eq:V_moments}) in Appendix~\ref{app:moments}.
\item[2.] Transform $V_i$ to the random variable $W_i$ with support $[-1,1]$ via (\ref{eq:W_transform}).
\item[3.] Compute the basis expansion coefficients $a_n$ via (\ref{eq:an_ev}).
\item[4.] Compute the basis expansion of $f_{W_i}(x)$ using (\ref{eq:W_exp}).
\item[5.] Transform $f_{W_i}(x)$ via (\ref{eq:VW_pdf_trans}) to obtain $f_{V_i}(x)$.
\end{algorithmic}
\end{algorithm}

\subsection{Recursion for the Net Profit PDF}\label{sec:panjer_rec}

The second stage is to compute the PDF of the net profit. Recall from Definition~\ref{def:rev_surp}, that the net profit for time slot $i$ is given by
\begin{align}
S_{net}(i) = \sum_{j=1}^{N_i}\left[v(\tau_{i,j},\mathbf{c}_{i,j}) - C(\tau_{i,j},k_{i,j})\right].
\end{align}
It is important to note that $S_{net}(i)$ has a \textit{different} distribution for each $i$ (see Remark~\ref{rem:dist_vary}). Moreover, observe that the distribution of $C(\tau_{i,j},k_{i,j})$ is discrete and the approximation of the distribution of $v(\tau_{i,j},\mathbf{c}_{i,j})$ from Section~\ref{sec:orthog_rep} is continuous (with bounded support). Also note that $S_{net}(i)$ is a random sum (in both the summands and number of summands). As such, it is not straightforward to efficiently compute the distribution of $S_{net}(i)$.

Fortunately, linear recursions have been developed to evaluate distributions for sums closely related to $S_{net}(i)$. In order to apply these recursive approaches, we first need to find the distribution of $v(\tau_{i,j},\mathbf{c}_{i,j})$ after it has been discretized. The discretization of $v(\tau_{i,j},\mathbf{c}_{i,j})$ can be obtained via standard approaches such as the Lloyd algorithm \cite{Gersho1992}; although it is important to carefully consider any negative terms in the distribution of $v(\tau_{i,j},\mathbf{c}_{i,j})$ that might occur due to the basis expansion approximation. We denote the discretization as $\hat{v}(\tau_{i,j},\mathbf{c}_{i,j})$.

To obtain the PDF of the net profit, denoted by $h_{Z_j}$, we now embed the discretized income into the lattice $\Delta\mathbb{Z}$ by choosing $\Delta > 0$ sufficiently small and rounding the support of $\hat{v}(\tau_{i,j},\mathbf{c}_{i,j})$. We then obtain the distribution
\begin{align}
h_{Z_j}(k) = \mathrm{Pr}(Z_j = k),
\end{align}
of the random quantity
\begin{align}
Z_j = \left[\hat{v}(\tau_{i,j},\mathbf{c}_{i,j}) - C(\tau_{i,j},k_{i,j})\right],
\end{align}
by convolving the (discrete) distributions of $\hat{V} =  \hat{v}(\tau_{i,j},\mathbf{c}_{i,j})$ and $C(\tau_{i,j},k_{i,j})$, which both have discrete support on $S \subset \Delta\mathbb{Z}$, with $|S| < \infty$.

At this point, we obtain the approximate distribution of $S_{net}(i)$. This is based on the following linear recurrence based on \cite[Eq. (1.7)]{Hurlimann1991},
\begin{align}\label{eq:f_dist}
&f_{S_{net}(i)}((n+1)\Delta)\notag\\
&~~= \frac{w_N}{1 - w_Nh_{Z_i}(0)}\notag\\
&~~~~\times\sum_{k=-\infty,~k\neq -1}^\infty (n+1)h_{Z_i}((k+1)\Delta)f_{S_{net}(i)}((n-k)\Delta).
\end{align}
We note that similar difference equations can be derived for distributions of $N_i$, other than geometric; in fact, it is straightforward to extend to any distribution in the Katz family (i.e., binomial, Poisson, and negative binomial distributions). Further details may be found in \cite[Eq. (1.3)]{Hurlimann1991} and \cite[Eq. (1)]{Panjer1981}.

\begin{remark}
We emphasize that the support of $S_{net}(i)$ is over $\mathbb{Z}$, not just $\mathbb{N}$. This means that the standard recursion in \cite{Panjer1981} (known as the Panjer recursion) is not applicable; the more general approach in \cite{Hurlimann1991} is required.
\end{remark}

In practice, the sum (\ref{eq:f_dist}) is in fact bounded (due to the finite resolution of the discretization step), which gives
\begin{align}\label{eq:f_dist_approx}
&f_{S_{net}(i)}((n+1)\Delta)\notag\\
&~~= \frac{w_N}{1 - w_Nh_{Z_i}(0)}\notag\\
&~~~~\times\sum_{k=k_{\min},~k\neq -1}^{k_{\max}} (n+1)h_{Z_i}((k+1)\Delta)f_{S_{net}(i)}((n-k)\Delta),
\end{align}
where\footnote[3]{The set $\mathrm{Supp}(S_{net}(i))$ corresponds to the support of $S_{net}(i)$.} $k_{\min} = \min \mathrm{Supp}(S_{net}(i)) - 1$ and $k_{\max} = \max \mathrm{Supp}(S_{net}(i)) - 1$, depend on the discretization resolution.

In order to solve the linear recurrence relation in (\ref{eq:f_dist_approx}), we use a convex reformulation in terms of a least squares problem,
\begin{align}\label{eq:num_ls}
\min_{\mathbf{f}:~\mathbf{f}^T\mathbf{f} = 1} \|\mathbf{A}\mathbf{f}\|_2,
\end{align}
where $\mathbf{A}$ consists of the coefficients in terms of $h_{Z_i}$ from (\ref{eq:f_dist_approx}) and $\mathbf{f}$ is the distribution of $S_{net}(i)$. We note that in principle other objectives can be used in (\ref{eq:num_ls}), such as those discussed in \cite{Hurlimann1991}; however, we have found via numerical experiments that the least squares objective is suitable for the purpose of approximating the ruin probability.

\subsection{Recursion for the Probability of Ruin}\label{sec:ruin_rec}

The third (and final) stage of the calculation is to compute the probability of ruin. This is achieved by collecting the distributions of $S_{net}(l)$ and substituting into the new ruin probability recursion, which we derive in this section.

It is important to note that the ruin probability (defined in Definition~\ref{def:ruin_prob}) is the probability that the revenue surplus is negative at \textit{any} $l \leq L$. Formally, the ruin probability can be written as
\begin{align}
\psi_L(u) = \mathrm{Pr}(S_{1}(u) < 0 \cup \cdots \cup S_{L}(u) < 0).
\end{align}
It is helpful to write the ruin probability in terms of the survival probability (see Definition~\ref{def:ruin_prob}), which is given by
\begin{align}\label{eq:surv_prob_exp}
\phi_L(u) &= 1 - \psi_L(u) = \mathrm{Pr}(S_{1} \geq 0 \cap \cdots \cap S_{L} \geq 0).
\end{align}

We now give the recursion for the probability of ruin in terms of the distributions of $S_{net}(l),~l = 1,2,\ldots,L$. First, define
\begin{align}
G_l(y) = \mathrm{Pr}(S_{net}(l) \leq y).
\end{align}
We then have the following theorem, which generalizes the ruin probability recursion in \cite{Sun2003,DeVylder1988}.
\begin{theorem}\label{thrm:ruin_rec}
The survival probability after $L$ time periods (e.g., months) is given by
\begin{align}
\phi_L(u) &= \phi_{L-1}(u)(1 - G_L(0))\notag\\
&~~~+ \int_{-\infty}^{0} \phi_{L-1}\left(u + \frac{y}{(1 + r)^L}\right)dG_L(y), \;\; \forall L \geq 2,
\end{align}
where
\begin{align}
\phi_1(u) = 1 - G_1(-u(1 + r)).
\end{align}
The ruin probability is then obtained via $\psi_L(u) = 1 - \phi_L(u)$.
\end{theorem}
\begin{IEEEproof}
See Appendix~\ref{app:ruin_rec}.
\end{IEEEproof}

We note that the integral in Theorem~\ref{thrm:ruin_rec} is a Stieltjes integral. This is important as $G_l,~l = 1,2,\ldots,L$ is in fact a sequence of distributions, all with discrete and bounded support. As such, the Stieltjes integral is a finite sum and can be efficiently evaluated numerically.

\section{Numerical and Simulation Results}\label{sec:sim_results}

With our evaluation framework based on ruin-theory in hand, we now turn to evaluating facility network sharing. We obtain pricing schemes that yield a probability of ruin less than $10\%$, under practical operating conditions. Along the way, we present important tradeoffs between physical layer resources such as pathloss, and financial constraints such as initial capital, interest rate (compounded monthly) and pricing.

Before evaluating the ruin probability using the techniques we have developed in Section~\ref{sec:eval}, we first present a simplified analysis based purely on the expected revenue surplus (see Definition~\ref{def:rev_surp}). This analysis provides initial insights into the financial aspects of the problem, which may be unfamiliar to the wireless communication community. That is, we consider
\begin{align}
\mathbb{E}[S_{net}(l)] = u(1 + r)^l + \sum_{i = 1}^l (1 + r)^{l - i} \mathbb{E}[N]\left(\mathbb{E}[V] - \mathbb{E}[C]\right),
\end{align}
where we assume that the connection duration time $\tau_{i,j}$ is constant, which means that the revenue per user $V_i$ has expectation $\mathbb{E}[V]$ for all $i$. The condition on the initial capital to ensure positive average revenue surplus (i.e., $\mathbb{E}[S_{net}(l)] \geq 0$) at a given time $n$ for interest rate $r$ is then given by
\begin{align}\label{eq:uBound}
u \geq \mathbb{E}[N]\left(\mathbb{E}[C] - \mathbb{E}[V]\right)\left(\frac{(1 + r)^n - 1}{r(1 + r)^n}\right).
\end{align}

\begin{figure}[!h]
\centerline{\includegraphics[height=3in]{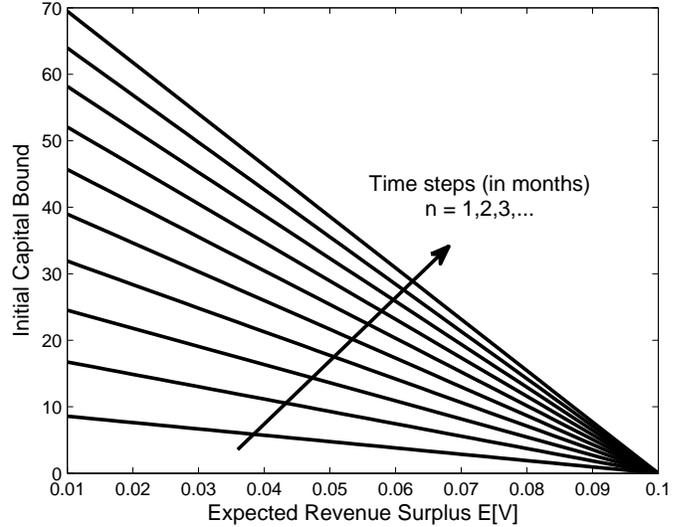}}
\caption{Plot of bound on initial capital $u$ to ensure $\mathbb{E}[S_{net}(l)] \geq 0$, where $r = 0.05,\mathbb{E}[N] = 100,\mathbb{E}[C] = 0.1$.} \label{fig:exp_surplus_init_capital}
\end{figure}

To illustrate this condition, Fig.~\ref{fig:exp_surplus_init_capital} plots the expected revenue $\mathbb{E}[V]$ versus the bound on $u$; each curve corresponding to a different time slot $n$. First, observe the initial capital needs to be non-zero to ensure a positive \textit{expected revenue surplus} when the costs exceed the revenue; i.e., $\mathbb{E}[C] > \mathbb{E}[V]$. Second, observe that there is a diminishing increase in the required initial capital as the number of time steps is increased; this can be seen by comparing the gaps between the curves, for a fixed $\mathbb{E}[V]$. Moreover, as the bound on $u$ in (\ref{eq:uBound}) scales linearly with $\mathbb{E}[N]$, the initial capital increases significantly as the number of connected users increase, when $\mathbb{E}[C] > \mathbb{E}[V]$. These are general trends; however, it is important to note the limitations of any analysis based on \textit{expected} revenue surplus. In particular, short-term behavior of the revenue surplus is not accounted for, which has the consequence that fluctuations in user demand or available network resources can cause the revenue surplus to drop below zero---leading to ruin. As such, it is necessary to use our framework based on ruin theory to account for these fluctuations and ultimately reduce the financial risk to the facility.

To account for fluctuations in the revenue surplus, we now turn to our evaluation framework based on ruin theory. We consider the network setup in Table~\ref{table:network_setup} and define $A_d = 2^{R^{(1)}/B} - 1$, where $R^{(1)}$ is the rate product on offer.

\begin{table}[!h]
\begin{center}\caption{Summary of network parameters.}\label{table:network_setup}
 \begin{tabular}{|c|c|c|c|c|c|c|c|c|}
 \hline
 Parameter & $K$ & $\sigma^2$ &  $P_0 = P_I$ & $w_N$ & $Q$ & $T\rho$ & $C$ & $\kappa T$\\ \hline
 Value     & $1$ & $0$ & $1$ & $0.2$ & $1$ & $1$ & $100$ & $1$ month \\
 \hline
 \end{tabular}
\end{center}
\end{table}

\begin{table*}[!t]
\begin{center}\caption{Table showing effect of varying small-cell density $\beta$, for different pathloss exponent $\alpha$ using numerical (Num.) and Monte Carlo (M.C.) approaches. Evaluation performed with $c_{\min} = 0.1$, $c_{\max} = 100$, $A_d = 100$.}\label{table:effect_density}
  \begin{tabular}{|c|c|c|c|c|}
    \hline
    $\beta$ & $\mathbb{E}[V]$ Num., $\alpha = 3$ & $\mathbb{E}[V]$ M.C., $\alpha = 3$ & $\mathbb{E}[V]$ Num., $\alpha = 4$ & $\mathbb{E}[V]$ M.C., $\alpha = 4$ \\
    \hline
    \hline
    0.01 & 76.5 & 75.5 & 60.5 & 60.2 \\
    \hline
    0.1 & 76.5 & 77.0 & 60.5 & 60.9 \\
    \hline
    1 & 76.5 & 76.4 & 60.5 & 59.8 \\
    \hline
  \end{tabular}
\end{center}
\end{table*}

We first consider the moments of the revenue $\mathbb{E}[V]$. These moments are ultimately used to compute the probability of ruin. They are also of interest in their own right. We show this next by examining the role of the key wireless network and financial parameters $\alpha$, $\beta$, $c_{\min}$, and $c_{\max}$.

In Table~\ref{table:effect_density}, we compare the small-cell density $\beta$ with the first moment of $V$, $\mathbb{E}[V]$, obtained numerically via (\ref{eq:V_moments}) and via Monte Carlo simulation. We point out that the moments obtained numerically via (\ref{eq:V_moments}) are in good agreement with the moments obtained via Monte Carlo simulation. Importantly, the moment $\mathbb{E}[V]$ is constant irrespective of the small-cell density for both the numerical and Monte Carlo approaches. This suggests that the small-cell density does not play an important role in networks well-modeled by PPPs. We believe the reason for this is that as the density of small-cells increases, there is an increase in nearby interfering small-cells, which is balanced by a closer servicing small-cell. Note that the small-cell density has a similar effect of outage probability in the low noise region, as shown in \cite{Andrews2011}.

\begin{figure}[!h]
\centerline{\includegraphics[height=3in]{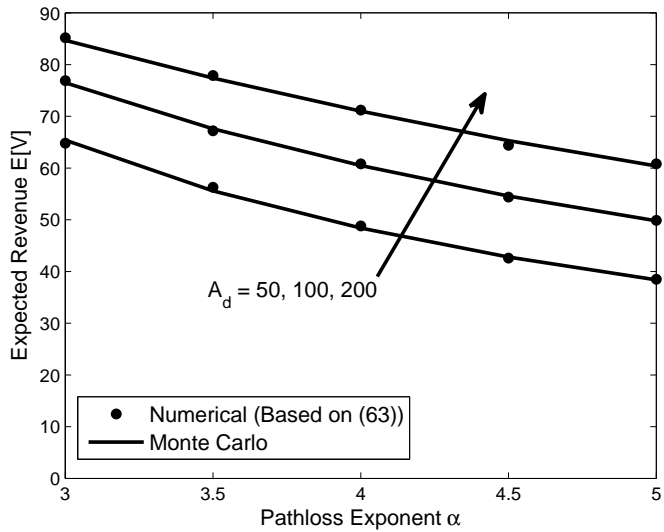}}
\caption{Plot of expected revenue $\mathbb{E}[V]$ versus pathloss exponent $\alpha$, with varying $A_d = 2^{R^{(1)}/B} - 1$. Parameters are $\beta = 0.1$, $c_{\min} = 0.1$, $c_{\max} = 100$, $A_d = 100$.} \label{fig:mom_alpha}
\end{figure}

Next, we consider the effect of the pathloss on the revenue. Fig.~\ref{fig:mom_alpha} plots the pathloss exponent versus the expected revenue $\mathbb{E}[V]$, which shows excellent agreement between our numerical result (based on (\ref{eq:V_moments})) and Monte Carlo results. The trend illustrated by the figure is that the revenue decreases as the pathloss exponent increases, irrespective of $A_d$ (corresponding to different rate products $R^{(1)}$ on offer). This is due to the fact that it is easier to service users with a high pathloss exponent as it means that there are often lower interference levels.

\begin{figure}[!h]
\centerline{\includegraphics[height=3in]{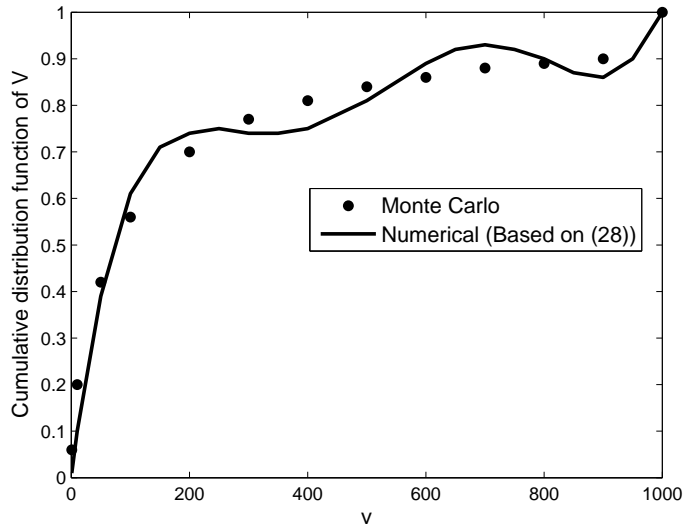}}
\caption{Plot of basis expansion approximation for the CDF of $V$ using four moments. Parameters are $\alpha = 4$, $\beta = 0.1$, $c_{\min} = 0.001$, $c_{\max} = 1000$, $A_d = 100$.} \label{fig:basis_expansion}
\end{figure}

Following Algorithm~\ref{alg:Ruin_Prob_Calc}, the next step to compute the ruin probability is to obtain the PDF of the revenue $V$ via basis expansion. To illustrate this step, the basis expansion approximation for the CDF of $V$ is plotted in Fig.~\ref{fig:basis_expansion}. We observe that our numerical approximation is in good agreement for revenue less than $200$, and then has small oscillations for high revenue values. It is important to note that tail oscillations are common when using the Jacobi polynomial representation, and must be carefully accounted for. In the setup for Fig.~\ref{fig:basis_expansion}, the approximation is good; however, as $c_{\min}$ is increased or $c_{\max}$ decreased, the discontinuity affects the quality of the approximation. As such, additional moments (i.e., $\mathbb{E}[V^5]$,..) are required to smooth the oscillations, which can be obtained easily from our analysis in Appendix~\ref{app:moments}. We also note that the parameters $a,b$ in the Jacobi polynomials (see Section~\ref{sec:orthog_rep}) strongly influence the approximation; extensive numerical experiments suggest that $a = b = 1$ is a robust choice.

Table~\ref{table:ruin} shows the ruin probability after $5$ months versus the initial capital $u$, for varying $A_d$ (reflecting different rate products on offer). As expected from Fig.~\ref{fig:exp_surplus_init_capital}, in order to obtain a low probability of ruin, the initial capital needs to be chosen carefully. This is reflected in both the numerical and Monte Carlo results. We note that for low ($u = 100$) and high ($u \geq 250$) initial capital, the numerical and Monte Carlo approaches are in agreement. However, for $u \approx 200$, there is a difference of approximately $0.1$. This is due to the discretization step detailed in Section~\ref{sec:panjer_rec}. Importantly, to ensure the probability of ruin is less than approximately $10\%$, an initial capital of $u > 250$ is required, with the parameters used in Table~\ref{table:ruin}.

\begin{table}[!h]
\begin{center}\caption{Ruin probability after $5$ months with varying initial capital $u$. Parameters are $\alpha = 4$, $\beta = 0.1$, $c_{\min} = 0.001$, $c_{\max} = 1000$, $A_d = 100$, and $r = 0.05$. }\label{table:ruin}
 \begin{tabular}{|c|c|c|c|c|c|}
 \hline
 Initial Capital $u$ & $100$ & $150$ &  $200$ & $250$ & $300$\\ \hline
 Numerical (Using Theorem~\ref{thrm:ruin_rec}) & $0.33$ & $0.09$ & $0.01$ & $0$ & $0$ \\ \hline
 Monte Carlo & $0.33$ & $0.22$ & $0.11$ & $0.05$ & $0.02$ \\ \hline
 \end{tabular}
\end{center}
\end{table}

\section{Conclusions}\label{sec:conc}

Due to the recent availability of cheap small-cells and the unique operating requirements of facilities, there is a need for alternative network sharing arrangements. To this end, we proposed a facility network sharing arrangement, which is based on: a leasing agreement between traditional operators and the facility; and a service agreement between users and the facility. Unlike traditional operators, the local scale of facilities means that technical design at the level of the network architecture is intimately connected with the profitability; instead of loosely coupled.

In order to evaluate our facility network sharing arrangement, we adopted a socio-technical system design approach. As such, our key performance metric is the ruin probability---leading to a new ruin-based evaluation framework. To evaluate the ruin probability, we proposed a numerical approximation method, which is shown to be in good agreement with Monte Carlo simulation. Our approximation method includes two key novel aspects: computation of the moments of the revenue via stochastic geometry techniques; and a new recursion for the ruin probability, which is tailored for the wireless communication setting. Our numerical results suggested that there are in fact concrete conditions where profitable operation of facilities is possible, with sufficient initial capital.

\section*{Acknowledgements}

This publication was supported by the European social fund within the framework of realizing the project ``Support of inter-sectoral mobility and quality enhancement of research teams at Czech Technical University in Prague'', CZ.1.07/2.3.00/30.0034. Period of the project´s realization 1.12.2012 – 30.6.201.

\appendices

\section{}\label{app:moments}

In this appendix, we derive the moments of the income for a user with connection ending in time slot $i$, which is used to compute the ruin probability. Importantly, our analysis can be straightforwardly extended to higher moments. Recall that the income from a single user (a typical node, by Slivnyak's theorem for homogeneous Poisson point processes) is given by
\begin{align}
v(\tau_{i,j},\mathbf{c}_{i,j}) = \sum_{l=1}^{\tau_{i,j}} c_{i,j,l}T\rho,
\end{align}
where the scaling factor is
\begin{align}
c_{i,j,l} = \max\left\{\min\left\{\frac{\left(2^{R_{i,j}/B} - 1\right)P_II_{i,j,l}}{|h_{i,j,l}|^2d_{i,j}^{-\alpha}P_0},c_{\max}\right\},c_{\min}\right\}.
\end{align}
We also define: $A = \frac{P_I(2^{R_{i,j}/B} - 1)}{P_0d_{i,j}^{-\alpha}}$; $H_l = |h_{i,j,l}|^2$; $\tau = \tau_{i,j}$; and $V = v(\tau_{i,j},\mathbf{c}_{i,j})$.

In order to compute the four moments, we require the moment generating function (MGF) of $V$. In turn, the MGF is obtained from the Laplace transform $\mathcal{L}_{X}(t)$, which is given by
{\small{
\begin{align}
\mathcal{L}_V(t) &= \mathbb{E}\left[\exp\left(-t\sum_{l=1}^{\tau} \max\left\{\min\left\{\frac{AI_l}{H_l},c_{\max}\right\},c_{\min}\right\}T\rho\right)\right]\notag\\
&= \mathbb{E}_{\tau}\biggl[\mathbb{E}_{A}\biggl[\biggl(\mathbb{E}_{H_l,I_l}\biggl[\exp\biggl(-t\biggr.\biggr.\biggr.\biggr.\biggr.\notag\\
&~~~\times \left.\left.\left.\left.\left.\max\left\{\min\left\{\frac{AI_l}{H_l},c_{\max}\right\},c_{\min}\right\}T\rho\right)\biggr\vert A\right]\right)^{\tau}\biggr \vert \tau\right]\right].
\end{align}}}
Note that $R_{i,j}$ and $\tau$ have discrete support, so these expectations are sums we evaluate last. We now evaluate the inner expectation over the interference $I$ and the channel gain $H_l$. Observe that the inner expectation can be written as
\begin{align}
E(t) &= \mathbb{E}_{H_l,I_l}\left[\exp\left(-t\max\left\{\min\left\{\frac{AI_l}{H_l},c_{\max}\right\},c_{\min}\right\}\right.\right.\notag\\
&~~~\times \left.\left.T\rho\right)\biggr\vert A,\tau\right]\notag\\
&= E_1(t) + E_2(t) + E_3(t),
\end{align}
where
\begin{align}
E_1(t) &= \mathbb{E}\left[e^{-tT\rho c_{\min}} \biggr\vert \frac{AI_l}{H_l} < c_{\min}\right]\mathrm{Pr}\left(\frac{AI_l}{H_l} < c_{\min}\right)\notag\\
&= e^{-tT\rho c_{\min}} \mathrm{Pr}\left(\frac{AI_l}{H_l} < c_{\min}\right),\notag\\
E_2(t) &= e^{-tT\rho c_{\max}} \mathrm{Pr}\left(\frac{AI_l}{H_l} > c_{\max}\right),\notag\\
E_3(t) &= \mathbb{E}\left[e^{-tT\rho \frac{AI_l}{H_l}} \biggr\vert c_{\min} \leq \frac{AI_l}{H_l} \leq c_{\max}\right]\notag\\
&~~~\times\mathrm{Pr}\left(c_{\min} \leq \frac{AI_l}{H_l} \leq c_{\max}\right)\notag\\
&= \int_{c_{\min}}^{c_{\max}} e^{-tT\rho w}f_W(w)dw,
\end{align}
with
\begin{align}\label{eq:W_def}
W = \frac{AI_l}{H_l}.
\end{align}
To compute $E_3(t)$, observe that
\begin{align}
\mathrm{Pr}(W \leq w) = \mathbb{E}_{I_l}\left[e^{-\frac{AI_l}{w}}\right] \Rightarrow ~f_W(w) = \mathbb{E}_{I_l}\left[\frac{AI_l}{w^2}e^{-\frac{AI_l}{w}}\right].
\end{align}
Hence,
\begin{align}
E_3(t) = \mathbb{E}_{I_l}\left[\int_{c_{\min}}^{c_{\max}} e^{-tT\rho w} \frac{AI_l}{w^2}e^{-\frac{AI_l}{w}}dw\right].
\end{align}
Next, we integrate by parts to obtain
\begin{align}\label{eq:et}
E(t) &= E_1(t) + E_2(t) + E_3(t)\notag\\
&= e^{-tT\rho c_{\max}} + tT\rho \int_{\frac{1}{c_{\max}}}^{\frac{1}{c_{\min}}} \frac{1}{u^2}e^{-tT\rho/u}\mathbb{E}_{I_l}\left[e^{-AI_lu}\right]du.
\end{align}

In order to compute $E(t)$, we require $\mathbb{E}_I[e^{-AI_lu}]$, which is given by
\begin{align}
\mathbb{E}_{I_l} \left[e^{-AI_lu}\right] &\overset{(a)}{=} \mathbb{E}_{\mathcal{X},\{g_m\}}\left[\prod_{m \in \mathcal{X} \setminus \{b_0\}} e^{-Au g_mr_m^{-\alpha}}\right]\notag\\
& \overset{(b)}{=} \mathbb{E}_{\mathcal{X}}\left[\prod_{m \in \mathcal{X} \setminus \{b_0\}} \mathbb{E}_g\left[e^{-Au gr_m^{-\alpha}}\right]\right]\notag\\
& \overset{(c)}{=} \exp\left(-2\pi \beta \int_{r_U}^{\infty} \left(1 - \mathbb{E}_g\left[e^{-Au gz^{-\alpha}}\right]\right)zdz\right),
\end{align}
where: $(a)$ follows from $H_l \sim \exp(1)$; $(b)$ follows from the fact that $\{g_m\}$ is independent of the spatial point process; and $(c)$ follows from the probability generating functional of the PPP\footnote[4]{The probability generating functional property of the PPP gives $\mathbb{E}[\prod_{x \in \mathcal{X}}f(x)] = e^{-\beta \int_{\mathbb{R}^2} (1 - f(x))dx}$.}.

Continuing, we have
\begin{align}\label{eq:I_ltemp}
&\mathbb{E}_{I_l} \left[e^{-AI_lu}\right]\notag\\
&~~= \exp\left(-2\pi\beta \int_0^\infty \frac{1}{\alpha}\int_0^{r_U^{-\alpha}} \left(1 - e^{-Agyu}\right)y^{-\frac{2}{\alpha} - 1}dy e^{-g}dg\right),
\end{align}
which follows from the change of variables $y = z^{-\alpha} \Rightarrow z = y^{-1/\alpha} \Rightarrow dz = -\frac{1}{\alpha}y^{-\frac{1}{\alpha} - 1}dy$.

Now consider the inner integral in (\ref{eq:I_ltemp}), given by
\begin{align}
F &= \int_0^{r_U^{-\alpha}} \left(1 - e^{-Agyu}\right)y^{-\frac{2}{\alpha} - 1}dy\notag\\
  &= -\frac{\alpha}{2}\left(1 - e^{-Agr_U^{-\alpha}u}\right)r_U^2 + \frac{\alpha A gu}{2}F_1,
\end{align}
where
\begin{align}
F_1 &= \int_0^{r_U^{-\alpha}} e^{-Agyu}y^{-\frac{2}{\alpha}}dy.
\end{align}
Integrating by parts again we obtain,
\begin{align}
F_1 &= \left(\frac{1}{Agu}\right)^{-\frac{2}{\alpha}+1} \gamma\left(1 - \frac{2}{\alpha},Agur_U^{-\alpha}\right),
\end{align}
where $\gamma(s,x) = \int_0^x t^{s-1}e^{-t}dt$.

Next, we compute the outer integral in (\ref{eq:I_ltemp}), which is given by
\begin{align}
G = \int_0^\infty Fe^{-g}dg = G_1 + G_2,
\end{align}
where
\begin{align}
G_1 &= -\frac{\alpha r_U^2}{2} \int_0^\infty \left(1 - e^{-Agur_U^{-\alpha}}\right)e^{-g}dg,\notag\\
G_2 &= \frac{\alpha}{2}\left(\frac{1}{Au}\right)^{-\frac{2}{\alpha}}\int_0^\infty g^{\frac{2}{\alpha}}e^{-g}\gamma\left(1 - \frac{2}{\alpha},Agur_U^{-\alpha}\right)dg.
\end{align}
We now require the identity from \cite[Eq. 6.4552]{Gradshteyn2007},
\begin{align}
\int_0^\infty x^{\mu - 1}e^{-\beta x}\gamma(\nu,\alpha x)dx  &= \frac{\alpha^{\nu}\Gamma(\mu + \nu)}{\nu(\alpha + \beta)^{\mu + \nu}}\notag\\
 &~~~\times{}_2F_1\left(1, \mu + \nu; \nu + 1; \frac{\alpha}{\alpha + \beta}\right).
\end{align}
Identifying terms, we obtain
\begin{align}
G_2 &= \frac{\alpha}{2}\left(\frac{1}{Au}\right)^{-\frac{2}{\alpha}}\frac{\left(Aur_U^{-\alpha}\right)^{1 - \frac{2}{\alpha}}}{\left(1 - \frac{2}{\alpha}\right)\left(Aur_U^{-\alpha} + 1\right)^2}\notag\\
&~~~\times{}_2F_1\left(1,2;2 - \frac{2}{\alpha};\frac{Aur_U^{-\alpha}}{1 + Aur_U^{-\alpha}}\right).
\end{align}

To compute $\mathbb{E}[e^{-AI_lu}]$, substitute $G_1$ and $G_2$ into (\ref{eq:I_ltemp}). This result is then used to compute the Laplace transform and obtain the moments via
\begin{align}\label{eq:moments}
\mathbb{E}[V_{i}^n] &= (-1)^n\frac{d^n\mathcal{L}_{V_i}(t)}{dt^n}|_{t=0}.
\end{align}

All that remains is to explicitly compute the moments. We define the following terms: $E_0^{(k)}(t)$ is the $k$-the derivative of $E(t)$; $E_0^{(k)}$ is the $k$-the derivative of $E(t)$ evaluated at $t = 0$; and $L_0^{(k)}$ is the $k$-th derivative of $\mathcal{L}_X(t)$ conditioned on $d,R_{i,j},\tau$ evaluated at $t = 0$. Note that $E(0) = 1$.

Using (\ref{eq:et}), we have
\begin{align}
E(t)  &= e^{-tT\rho c_{\max}} + tT\rho \int_{\frac{1}{c_{\max}}}^{\frac{1}{c_{\min}}} \frac{1}{u^2}e^{-tT\rho/u}\mathbb{E}_{I_l}\left[e^{-AI_lu}\right]du,\notag\\
E^{(s)}(t) &= (-T\rho c_{\max})^se^{-tT\rho c_{\max}}\notag\\
&~~~+ s(-1)^{s+1}(T\rho)^s\int_{\frac{1}{c_{\max}}}^{\frac{1}{c_{\min}}} \frac{1}{u^{s+1}}e^{-tT\rho/u}\mathbb{E}_{I_l}\left[e^{-AI_lu}\right]du\notag\\
&~~~-t(T\rho)^{s+1}\int_{\frac{1}{c_{\max}}}^{\frac{1}{c_{\min}}} \frac{1}{u^{s+2}}e^{-tT\rho/u}\mathbb{E}_{I_l}\left[e^{-AI_lu}\right]du,
\end{align}
where $E^{(s)}(t)$ is the $s$-th derivative of $E(t)$ and $\mathbb{E}[e^{-AI_lu}]$ is given by (\ref{eq:I_ltemp}).

Now, let $\mathbf{1}$ denote the indicator function. The moments conditioned on $d_{i,j},R_{i,j},\tau_{i,j}$ can now be readily obtained. We illustrate with the first conditional moment.
\begin{align}\label{eq:cond_moments}
L_0^{(1)} &= \tau E_0^{(1)}\mathbf{1}_{\tau > 0}.
\end{align}

Finally, the moments for the revenue from a user with connection ending in time slot $i$ are given by
\begin{align}\label{eq:V_moments}
\mathbb{E}[V_i^k] = \sum_{l = 1}^n\sum_{k = 1}^Q \mathrm{Pr}(R_{i,j} = k)\mathrm{Pr}(\tau = l)\int_0^\infty L_0^{(k)}e^{-\pi \beta z^2}2\pi \beta zdz,
\end{align}
which can be efficiently evaluated numerically.

\section{}\label{app:ruin_rec}

\begin{IEEEproof}[Proof of Theorem~\ref{thrm:ruin_rec}]
First (from (\ref{eq:surv_prob_exp})), observe that the survival probability can be written as
\begin{align}
\phi_n(u) &= \mathrm{Pr}\left(u(1 + r)^n + S_{net}(n) + \sum_{i=1}^{n-1} S_{net}(i)(1 + r)^{n-i} \geq 0, \right.\notag\\
&~~~\left. \ldots,u(1 + r) + S_{net}(1) \geq 0\right).
\end{align}
Also recall that
\begin{align}\label{eq:gn}
G_n(y) = \mathrm{Pr}(S_{net}(n) \leq y).
\end{align}
Now, using (\ref{eq:surv_prob_exp}) and (\ref{eq:gn}) yields $\phi_1(u) = \mathrm{Pr}(S_{net}(1) \geq -u(1 + r)) = 1 - G_1(-u(1+r))$. We then have for $n \geq 2$
\begin{align}
\phi_n(u) &= \mathrm{Pr}\left(u(1 + r)^{n-1} + \frac{S_{net}(n)}{1 + r}\right.\notag\\
&~~~\left. + \sum_{i=1}^{n-1} S_{net}(i)(1 + r)^{n-1-i} \geq 0,\ldots\right)\notag\\
&\overset{}{=} \phi_{n-1}(u)(1 - G_n(0))\notag\\
&~~~ + \int_{-\infty}^{0} \phi_{n-1}\left(u + \frac{y}{(1 + r)^n}\right)dG_n(y),
\end{align}
which follows after considering
\begin{align}
&\left\{u(1 + r)^{n-1} + \frac{S_{net}(n)}{1 + r} + \sum_{i=1}^{n-1} S_{net}(i)(1 + r)^{n-1-i} \geq 0\right.\notag\\
 &~~\left.\cap u(1 + r)^{n-1} + \sum_{i=1}^{n-1} S_{net}(i)(1 + r)^{n-1-i} \geq 0\right\}.
\end{align}
\end{IEEEproof}

\bibliographystyle{ieeetr}
\bibliography{Micronetworks}

\end{document}